\documentstyle[aps,manuscript,eqsecnum,prb]{revtex}

\begin{document}

\draft

\title{Barrier crossing induced by very slow external noise}

\author{Suman Kumar Banik, Jyotipratim Ray Chaudhuri 
and Deb Shankar Ray {\footnote{e-mail : pcdsr@mahendra.iacs.res.in}}
}

\address{Indian Association for the Cultivation of Science, Jadavpur,
Calcutta 700032, INDIA.}

\date{\today}

\maketitle

\begin{abstract}
We consider the motion of a particle in a force field subjected to adiabatic,
fluctuations of external origin. We do not put the restriction on the type
of stochastic process that the noise is Gaussian. Based on a method 
developed earlier by us [ J. Phys. A {\bf 31} (1998) 3937, 7301]
we have derived the equation of motion for probability distribution function 
for the particle on a coarse-grained timescale $\Delta t$ assuming that it 
satisfies the separation of timescales; $|\mu |^{-1} \ll \Delta t \ll \tau_c$, 
where $\tau_c$ is the correlation time of fluctuations. $|\mu |^{-1}$ refers 
to the inverse of the damping rate (or, the largest of the eigenvalues of the 
unperturbed system) and sets the shortest timescale in the dynamics in 
contrast to the conventional theory of fast fluctuations. The equation
includes a third order noise term. We solve the equation for a Kramers' type
potential and show that although the system is thermodynamically open, 
appropriate boundary conditions allow the distinct steady states. Based
on the exact solution of the third order equation for the linearized potential
and the condition for attainment of the steady states we calculate the
adiabatic noise-induced rate of escape of a particle confined in a well. A
typical variation of the escape rate as a function of dissipation which is
reminiscent of Kramers' turn-over problem, has been demonstrated.
\end{abstract}

\vspace{0.25cm}



\section{Introduction }

About a century ago Einstein formulated the problem of Brownian motion as what
is known today as stochastic processes. To explain \cite{furth,wax,kampen} the 
motion of a Brownian particle on an observable macroscopic scale he introduced 
the coarse-graining of space and time in the dynamics. This has 
served as the basis for generation of successive levels of description from 
the microscopic to macroscopic realm in subsequent many other formulations. 
Although there exists no general program of coarse-graining it has been possible to 
describe the dynamics fairly realistically in terms of some systematic 
separation of timescales consistent with experiments.

In dealing with the stochastic processes one essentially examines the average
motion of a system subjected to fluctuations which may be fast or slow
depending on the correlation time $\tau_c$ of the fluctuations compared to
coarse-grained time scale, $\Delta t$ over which one observes the 
{\it average motion} of the particle. While the stochastic processes with
short correlation time are well understood 
\cite{kampen,vanrep,uf1,uf2,hfa1,hfa2,ojh,cruty,rahman,cooper,asch1,asch2}, 
significant progress has been 
made in recent times in theories dealing with arbitrary long correlation time.
However, one important constraint in this context \cite{vanrep}
needs to be emphasized.
This is that the separation of timescales is just not enough and one has to
specify further the nature of stochastic process by assuming it to be 
either {\it Markovian} or {\it Gaussian} or {\it both} \cite{vanrep,mazo}. 
In fact these lie at the heart of the overwhelming majority of the traditional 
theories so far. The Ornstein-Uhlenbeck \cite{wax,ou1,ou2} process is an 
age old standard paradigm in this respect. 

Very recently we have proposed a method \cite{jpa1,jpa2}
for analyzing the stochastic dynamics of a particle subjected 
to {\it external}, adiabatically {\it slow} fluctuations where we 
do not limit ourselves to Markovian or Gaussian processes. It is apparent 
that to go beyond Markovian and Gaussian approximations the standard 
procedure of expansion of Master equation \cite{vanrep} (based on Markov
approximation) leading to Kramers-Moyal [KM] equation or characteristic 
function method \cite{mazo,adelman} (based on the calculation of moments and, 
in general, used for Gaussian processes) are not suitable for the purpose. 
Based on adiabatic following approximation we have derived the appropriate 
equations of motion for the
probability distribution function on a coarse-grained timescale $\Delta t$
assuming that $\Delta t$ satisfies
\begin{equation}
\label{ineqb}
\frac{1}{|\mu|} \ll \Delta t \ll \tau_c
\end{equation}

\noindent
where $\frac{1}{|\mu|}$ refers to the inverse of the damping rate (or the
inverse of the largest eigenvalue of the `unperturbed' system. This is in 
contrast to {\it fast} fluctuations characterized by the inequality
$\tau_c \ll \Delta t \ll \frac{1}{|\mu|}$).  $\frac{1}{|\mu|}$ therefore
refers to the shortest timescale in the dynamics in the present investigation
compared to short correlation time $\tau_c$ of the fast stochastic processes.
The separation of the timescales (\ref{ineqb}) has been critically analyzed
by Masoliver et. al. \cite{jaume1} in their treatment of generalized
Langevin equation for studying diffusion with external noise.

In this paper we extend our earlier treatment \cite{jpa1,jpa2} to analyze and 
present a solution for the problem of motion of a particle {\it in a force 
field} subjected to external, adiabatic noise pertaining to the separation of 
timescales (\ref{ineqb}). The potential we consider here is of Kramers' type as 
shown in Fig.(\ref{fig1}). The physical motivation behind the solution is two 
fold : 

First, since it is subjected to fluctuations of {\it external} origin, 
the system is thermodynamically open (the
close system on the other hand is characterized by internal noise which
satisfies the fluctuation-dissipation relation). We search for the condition
of an appropriate steady state for this open system. Similar studies for 
Gaussian processes have been carried out by Masoliver and others 
\cite{jaume1,jaume2}. The study of
open systems are specifically relevant for describing the effect of pump
fluctuations on the emission of a dye laser \cite{roy}, effect of 
fluctuating rate constants on a chemical reaction \cite{arnold,sancho} and
effect of noise on parametric oscillator \cite{kawa}, etc.

Second, once the condition for attainment of the steady state is realized, it
becomes possible to consider the situation such that the particle in a force
field, i.e., originally confined in a potential well may escape under the
influence of external adiabatic noise by maintaining a steady state 
probability current over the barrier. It is therefore pertinent to calculate
the rate of escape induced by this nonthermal activation in the spirit of 
Kramers and Smoluchowski \cite{kramers,rate,melnikov} and to elucidate the aspects 
of dependence of escape rate on dissipation. The counterpart of the latter 
issue in the theory of fast fluctuations is the well known turn-over problem
\cite{rate,melnikov}.

We thus intend to touch the three issues in the theory of {\it adiabatic}
fluctuations in the subsequent sections : First, based on the method of
`adiabatic following approximations' 
\cite{jpa1,jpa2,crisp,gris1,gris2,gris3,tla} together with a systematic
separation of timescales (\ref{ineqb}) to carry out an expansion in 
$|\mu|^{-1}$ as
developed by us in two earlier papers \cite{jpa1,jpa2} we obtain an equation of motion 
for probability distribution function for a particle in a force field
simultaneously subjected to external adiabatic noise. The correlations of
fluctuating forces give rise to second and third order diffusion
coefficients \cite{jpa2}. We discuss this issue
in Sec.~{II}. Second, although the system is thermodynamically open we look for
the physically allowed steady states and show that 
application of the appropriate boundary conditions leads to distinct 
steady states. Sec.~{III} is
devoted to this aspect. Third, based on the exact solution of third order
equation for the linearized potential and the condition for attainment of the 
steady states, we address the problem of escape of a particle confined in a 
well in the spirit of Kramers-Smoluchowski theory. This forms the subject
matter of Sec.~{IV}.  We conclude this paper by summarizing the main results and
their experimental relevance in Conclusion.


\section{ Motion of a particle in a force field in presence of 
adiabatic fluctuations }

The equation of motion of a particle of unit mass in an one-dimensional 
extension where it is acted upon by an external field of force corresponding 
to a potential $V(x)$ and an {\it external, adiabatic} stochastic force 
$\xi(t)$, can be written as follows
\begin{equation}
\dot{x} = -\frac{1}{\beta} V'(x) + \frac{1}{\beta} \; \alpha \;
\xi(t) \; \; .
\end{equation}

Here we have considered the overdamped limit. $\beta$ is the dissipation
constant and $\alpha$ which is necessary to keep track of the order of
perturbation, is a parameter determining the size of fluctuations of the 
external noise $\xi(t)$. We emphasize here the two points : (i) we do not
put restriction on the type of stochastic process $\xi(t)$ that
the noise is Gaussian. This has 
attracted so much attention in the recent literature that it is necessary
to point out that no such assumption has been made. The only restriction we 
make on the nature of the stochastic process $\xi(t)$ is that its correlation
time $\tau_c$ is very long, i.e., it corresponds to the separation of the 
timescales implied in the inequality (\ref{ineqb}). Also note that the 
inequality implies the overdamped limit.
(ii) We assume for convenience, without any loss of
generality, that $\langle\xi(t)\rangle=0$ \cite{kampen}.

In a preceding paper \cite{jpa2} we have derived the equation of motion for 
probability density distribution function $P(x,t)$ in phase space corresponding 
to the Langevin description (2) where the associated timescale satisfies the
inequality (\ref{ineqb}). We have shown that $P(x,t)$ obeys the differential
equation of motion which contains third order terms (beyond the usual 
Fokker-Planck terms) leading to non-Gaussian noise. The appearance of these
terms is generic for the stochastic process we consider here. The general
expression for time evolution of probability density function is given by
\cite{jpa2}
\begin{eqnarray}
\frac{\partial P(x,t)}{\partial t} & = & \left \{ -\nabla \cdot F_0 +
\alpha^2 \; \nabla \cdot \int_0^{\infty} \langle F_1 \nabla_{-\tau} \cdot
F_1(x^{-\tau}) \rangle \left |\frac{d x^{-\tau}}{dx} \right | \; d\tau
\right.  \nonumber \\
& & - \alpha^2 \; \nabla \cdot \int_0^{\infty} \tau \; \langle F_1 
\nabla_{-\tau} \cdot {\dot{F_1}} (x^{-\tau}) \rangle 
\left |\frac{d x^{-\tau}}{dx} \right | \; d\tau  \nonumber \\
& & \left. + \alpha^2 \; \nabla \cdot \int_0^{\infty} \tau \; \langle
F_1 \nabla_{-\tau} \cdot F_1 (x^{-\tau}) \; \nabla_{-\tau} \cdot F_0
(x^{-\tau}) \rangle \left |\frac{d x^{-\tau}}{dx} \right | \; d\tau \right
\} \; P(x,t) \; \; ,
\end{eqnarray}

\noindent
where $F_0$ and $F_1$ refer to the unperturbed and the fluctuating 
terms, respectively, corresponding to Eq.(2) as given by
\begin{equation}
F_0 = -\frac{1}{\beta} V'(x) \; \; \; \; {\rm and} \; \; \; \; 
F_1 = \frac{1}{\beta} \; \alpha \; \xi(t) \; \; .
\end{equation}

\noindent
The symbol $\nabla$ is used for the operator that differentiates everything
that comes after it with respect to $x$. $\nabla_{-\tau}$ denotes the 
differentiation with respect to $x_{-\tau}$. The Jacobian 
$\left |\frac{d x^{-\tau}}{dx} \right |$ defines the mapping 
$x\rightarrow x^{-\tau}$ for the unperturbed motion and is given by \cite{jpa2}
\begin{equation}
\left |\frac{d x^{-\tau}}{dx} \right | = 1 + \frac{1}{\beta} V''(x)\; \tau +
{\cal O}(\tau^2) \; \; ,
\end{equation}

\noindent
where it has been assumed that $x$ varies very little in $\tau$ (in the 
scale of $\frac{1}{\beta}$). Also explicitly we have
\begin{equation}
\nabla_{-\tau} = \left [ 1-\frac{1}{\beta} V''(x^{-\tau}) \; \tau \right ]
\frac{\partial}{\partial x} \; \; ,
\end{equation}
\begin{equation}
F_0 (x^{-\tau}) = -\frac{1}{\beta} V'(x^{-\tau}) = - \frac{1}{\beta} \left [
V'(x) - \tau \; V''(x) \right ]
\end{equation}
\noindent
and
\begin{equation}
F_1(x^{-\tau}) = \frac{1}{\beta} \xi(t-\tau) \; \; \; , \; \; \;
\dot{F}_1(x^{-\tau}) = \frac{1}{\beta} \left. \frac{d\xi(t)}{dt} \right
|_{(t-\tau)} \; \; .
\end{equation}

Making use of the relations (4-8) and after collecting the terms of the order
of $\alpha^2$, all the four terms in Eq.(3) can be simplified further to
obtain [ some details are outlined in the Appendix-A ] :

\begin{equation}
\label{fpeq}
\frac{\partial P(x,t)}{\partial t} = \frac{1}{\beta} 
\frac{\partial}{\partial x} \left [ V'(x) P(x,t) \right ] + 
\alpha^2\; c_{01} \; \frac{1}{\beta^2} \frac{\partial^2 P(x,t)}{\partial x^2} 
- \alpha^2 \; c_2 \; \frac{1}{\beta^3} \frac{\partial^3}{\partial x^3}
\left [ V'(x) P(x,t) \right ] \; .
\end{equation}

\noindent
$c_0$, $c_1$ and $c_2$ in Eq.(9) are given by
\begin{eqnarray}
c_{01} & = & c_0 - c_1 \nonumber \\
c_0 & = & \int_0^\infty \langle \xi(t) \; \xi(t-\tau) \rangle \; d\tau \nonumber \\
c_1 & = & \tau_c \langle \xi^2(t) \rangle - c_0 \nonumber \\
c_2 & =  &\int_0^\infty \tau \; \langle \xi(t) \; \xi(t-\tau) \rangle 
\; d\tau 
\; \; .
\end{eqnarray}

\noindent
We now put $\alpha=1$ for the rest of the treatment.

The above equation describes the time evolution of an overdamped particle
in a force field (derivable from a potential $V(x)$) simultaneously subjected
to an external adiabatic stochastic force. $c_0$, $c_1$ and $c_2$ measure
the strength of the noise term. While the first term in Eq.(9) can be identified
as the usual deterministic dynamical term, the second and the third terms refer
to second and third order diffusion coefficients due to stochasticity $\xi(t)$.
The remarkable departure from the standard form of Fokker-Planck equation
(Smoluchowski equation) is due to the presence of the {\it third order noise}.

Let us now digress a little bit about Eq.(\ref{fpeq}) which forms the basis of
this paper. The very appearance of third order term apparently suggests an
immediate bearing with the familiar KM expansion. We emphasize that 
Eq.(\ref{fpeq}) is not a truncated KM expansion. The KM expansion is based on 
an expansion of Master equation \cite{risken} in a series
in $\tau$ and the terms of ${\cal O}(\tau^2)$ are neglected. Or in other words
the coefficients in a KM expansion originate from the corresponding moments 
( of the
distribution function ) which are {\it assumed} to be linear in $\tau$. This
severely restricts the use of higher order moments for finite $\tau$. This
is mathematically formalized in terms of Pawula Theorem \cite{pawula}
( whose proof concerns the relation between the moments pertaining 
to the KM expansion only. Note that the familiar Wigner \cite{weq}
equation for probability distribution for a
cubic potential is a third order equation which has nothing to do with a KM
expansion or Pawula theorem ) which precludes the occurrence of nonzero terms
beyond Fokker-Planck in a KM expansion. Thus if there is any finite third
order term it has to be treated as a perturbation. In many situations one
encounters serious interpretive difficulties since the probability
distribution functions often turn out to be negative.

The expansion leading to Eq.(\ref{fpeq}) on the other hand is based on an
expansion [ Eq.(9) of \cite{jpa2} ] in $\alpha \beta^{-1}$ where $\alpha$ is the strength of noise and
$\beta$ is the damping constant. Note that $\beta^{-1}$ defines the shortest
time scale in the dynamics. That the expansion is in $\alpha \beta^{-1}$
is evident from the appearance of
$\frac{1}{\beta}$, $\frac{1}{\beta^2}$ and $\frac{1}{\beta^3}$ factors in
the successive terms in Eq.(\ref{fpeq}). Although the convergence of formal KM
expansion, in general, is not guaranteed ( it is built in from outside as
done by van Kampen \cite{kampen} in terms of $\Omega^{-1}$ expansion ) it has been proved
that \cite{jpa1} our expansion scheme is convergent in the adiabatic
following limit. The appearance of third derivative term is
generic since it appears in the same order ( $\alpha^2$-order ) as the second
derivative term in the present expansion scheme \cite{jpa2} and is
characteristic of non-Gaussian features. Thus the 
third order term cannot be treated as a perturbation.

Eq.(\ref{fpeq}) is derived \cite{jpa1,jpa2} on the basis of {\it two 
approximations} : (i)
{\it adiabatic following approximation} which results in a convergent
expansion in $\alpha \beta^{-1}$ ( valid for very slow noise processes
corresponding to the timescales (\ref{ineqb}) ) which is complementary to
the cummulant expansion based on an expansion in $\alpha \tau_c$ ( valid
for very fast noise processes \cite{vanrep} ).
(ii) {\it decoupling approximation}. We have calculated the error caused by
decoupling  and shown \cite{jpa1,jpa2} that it is  of the order of
$\alpha^2 \beta^{-1}$. The corresponding error in decoupling in the case of
fast fluctuations is $\alpha^2 \tau_c$. We explicitly point out that no other
approximation is required to develop the theory further as done in this paper
and given the appropriate boundary conditions the derived
solutions are well behaved positive definite probability distribution
functions.


\section{ The solution of third order equation for a Kramers' type
potential : Steady state probability density }

\subsection{The general solution}

To start with we now recast the third order equation (9) in the form of the
familiar continuity equation and identify the current $S(x,t)$ as follows;
\begin{equation}
\frac{\partial P(x,t)}{\partial t} = -\frac{\partial S(x,t)}{\partial x}
\end{equation}

\noindent
where
\begin{equation}
S(x,t) = - \frac{1}{\beta}\; V'(x) \; P - \frac{c_{01}}{\beta^2} \;
\frac{\partial P}{\partial x} + \frac{c_2}{\beta^3} \;
\frac{\partial^2}{\partial x^2} \left \{ V'(x) P \right \} \; \; .
\end{equation}

\noindent
In the steady state one puts
\begin{equation}
\frac{\partial P(x,t)}{\partial t} = 0
\end{equation}

\noindent
to obtain the following steady state current $J$ as given by
\begin{equation}
J = - \frac{1}{\beta}\; V'(x) \; P_s - \frac{c_{01}}{\beta^2} \;
\frac{\partial P_s}{\partial x} + \frac{c_2}{\beta^3} \;
\frac{\partial^2}{\partial x^2} \left \{ V'(x) P_s \right \} \; \; ,
\end{equation}

\noindent
where $P_s$ is the steady state probability distribution. Henceforth for the
sake of brevity we omit the subscript $s$ from $P_s$ for all the subsequent 
calculations to follow and denote $P(x)$ as the steady state probability
distribution function.

Multiplying both sides of Eq.(14) by $\beta^3/c_2$ we obtain
\begin{equation}
\frac{d^2}{dx^2} \left \{ V'(x) P \right \} -a\frac{dP}{dx} -b\left \{
V'(x) P\right \} = \frac{\beta^3 J}{c_2} \; \; ,
\end{equation}

\noindent
where $a$ and $b$ are given by
\begin{equation}
a = \frac{\beta c_{01}}{c_2} \; \; \; \; {\rm and} \; \; \; \;
b = \frac{\beta^2}{c_2} \; \; .
\end{equation}

\noindent
We now explicitly make use of the Kramers' type potential $V(x)$ as shown
in Fig.(\ref{fig1}) for the problem. We consider by linearizing it at $x=0$,
\begin{equation}
V(x) = E_0 -\frac{1}{2}\omega_0^2 \; x^2
\end{equation}

\noindent
where $\omega_0$ refers to frequency of the inverted well. 
$E_0$ defines the potential at the barrier top.

The probability distribution function $P(x)$ therefore satisfies the
following equation 
\begin{equation}
x\frac{d^2 P}{dx^2} + \left ( \frac {2\omega_0^2 +a}{\omega_0^2} \right )
\frac{dP}{dx} -b\; x\; P
= - \frac{\beta^3 J}{c_2\; \omega_0^2} \; \; .
\end{equation}

\noindent
Defining
\begin{equation}
\gamma  =   \frac {2\omega_0^2 +a}{\omega_0^2} \; \; \; {\rm and} \; \; \;
D = - \frac{\beta^3 J}{c_2\; \omega_0^2} \; \; ,
\end{equation}

\noindent
the Eq.(18) reads as 
\begin{equation}
x\frac{d^2 P}{dx^2} + \gamma \; \frac{dP}{dx} - b\; x\; P
= D \; \; .
\end{equation}

\noindent
It is convenient to make the following substitution 
\begin{equation}
P(x) = x^{\frac{1}{2}(1-\gamma)} \; W(x)
\end{equation}

\noindent
in Eq.(20) to obtain
\begin{equation}
x^2\frac{d^2 W}{dx^2} + x \; \frac{dW}{dx} - \left [\frac{1}{4} \left (
\gamma-1 \right )^2 + b\; x^2\right ] \; W
= D \; x^{\frac{1}{2}(1+\gamma)} \; \; .
\end{equation}

\noindent
Let
\begin{equation}
\nu = \frac{1}{2} (\gamma-1) \; \; .
\end{equation}

\noindent
We then rewrite Eq.(22) as follows,
\begin{equation}
x^2\frac{d^2 W}{dx^2} + x \; \frac{dW}{dx} - (\nu^2 + b\; x^2 ) \; W
= D \; x^{1+\nu} \; \; .
\end{equation}

\noindent
From the definitions of $\gamma$ and $a$ (Eqs.(19) and (23)) we have
\begin{eqnarray}
\nu & =&  \frac{1}{2} \left ( 1+\frac{a}{\omega_0^2} \right ) \nonumber \\
& = & \frac{1}{2} \left ( 1+ \frac{\beta c_{01}}{\omega_0^2 c_2} \right ) \; \; . 
\end{eqnarray}

\noindent
The structure of $c_{01}(=c_0-c_1)$ suggests (see the definition (10)) that
by virtue of adiabatic stochasticity, $c_1$ is much smaller compared to $c_0$
and $c_{01}$ is always positive. This ensures that $\nu$ as defined in Eq.(25)
is always positive. 

It is convenient to make a further substitution for independent variable 
$x$ as
\begin{equation}
\zeta = \sqrt{b} \;x \; \; .
\end{equation}

\noindent
This reduces Eq.(24) to
\begin{equation}
\label{mbf}
\zeta^2\frac{d^2 W}{d\zeta^2} + \zeta \; \frac{dW}{d\zeta} - (\nu^2 + \zeta^2 ) \; W
= \frac{D}{b^{\frac{1}{2}(1+\nu)}}  \; \zeta^{1+\nu} \; \; .
\end{equation}

\noindent
The homogenous counterpart corresponding to the above Eq.(\ref{mbf}) 
is the standard modified Bessel equation of order
$\nu$. The general solution of Eq.(\ref{mbf}) can be written as
\begin{eqnarray}
W(\zeta) & = &  A\; I_\nu(\zeta) + B\; K_\nu(\zeta) \nonumber \\
& & + \frac{D}{b^{\frac{1}{2}(1+\nu)}} \; I_\nu(\zeta) \int^\zeta \zeta^{\prime\nu}
\; K_\nu (\zeta') \; d\zeta' 
- \frac{D}{b^{\frac{1}{2}(1+\nu)}} \; K_\nu(\zeta) \int^\zeta \zeta^{\prime\nu}
\; I_\nu (\zeta') \; d\zeta' \; \; ,
\end{eqnarray}

\noindent
where $I_\nu(\zeta)$ and $K_\nu(\zeta)$ are modified Bessel functions of order
$\nu$; $A$ and $B$ are the two arbitrary constants of integration corresponding 
to the homogenous part of Eq.(\ref{mbf}). The $D$ containing term results from the 
particular integral corresponding to the inhomogenous contribution of Eq.(\ref{mbf}) 
obtained by the method of variation of parameters. 
Making use of the relations (26) and (21) we revert back to the original 
variables $x$ and $P(x)$ to obtain the general solution of Eq.(20) as
\begin{eqnarray}
P(x) & = &  A\; x^{-\nu}\; I_\nu(\sqrt{b} x) + B\; x^{-\nu}\; 
K_\nu(\sqrt{b} x) \nonumber \\
& & + D \; x^{-\nu}\; \left [ I_\nu(\sqrt{b} x) \int^{\sqrt{b}x} x^{\prime\nu}
\; K_\nu (\sqrt{b} x') \; dx' -
K_\nu(\sqrt{b} x) \int^{\sqrt{b}x} x^{\prime\nu}
\; I_\nu (\sqrt{b} x') \; dx'  \right ] \; \; .
\end{eqnarray}

\subsection{The boundary conditions and the normalized probability
distribution}

At this juncture it is necessary to specify the boundary conditions. We 
impose the following natural boundary conditions on the solution (29)

\noindent
(i) $P(x)$ vanishes for $|x|\rightarrow \infty$\\
(ii) $P(x)$ remains finite at $x=0$.

\noindent
To this end we proceed as follows : 

\noindent
(i) First we note that a modified Bessel equation of the form
\begin{equation}
x^2\; y'' + x\; y' - (x^2+\nu^2)\; y = 0
\end{equation}

\noindent
has an irregular singular point as $x\rightarrow\infty$. The leading
behavior of the solutions are \cite{bo}
\begin{mathletters}
\begin{equation}
y(x) = I_\nu(x) \sim {\cal C}_1 \; x^{-1/2} \; e^x \; \; , \; \; \; \; \; \; \; \;
x\rightarrow \infty
\end{equation}
and
\begin{equation}
y(x) = K_\nu(x) \sim {\cal C}_2 \; x^{-1/2} \; e^{-x} \; \; , \; \; \; \; \; \; \; \;
x\rightarrow \infty \; \; .
\end{equation}
\end{mathletters}

\noindent
We thus observe that $I_\nu(x)$ diverges exponentially for large $x$. By
applying the boundary condition (i), i.e., $P(x)$ vanishes for large $x$
we see that the constant $A$ in the general solution (29) must be zero.
Therefore we have
\begin{eqnarray}
\label{gsol}
P(x) & = & B\; x^{-\nu}\; K_\nu(\sqrt{b} x) 
+ D \; x^{-\nu}\; I_\nu(\sqrt{b} x) \int^{\sqrt{b}x} x^{\prime\nu}
\; K_\nu (\sqrt{b} x') \; dx' \nonumber \\
& & - D \; x^{-\nu}\; K_\nu(\sqrt{b} x) \int^{\sqrt{b}x} x^{\prime\nu}
\; I_\nu (\sqrt{b} x') \; dx'  \; \; .
\end{eqnarray}

\noindent
$B$ and $D$ are the two remaining constants to be determined.

\noindent
(ii) We now turn to the second boundary condition, i.e., the finiteness of
$P(x)$ at $x=0$. For a fixed $\nu$ $(\nu>0)$ we know that \cite{bo}
\begin{equation}
K_\nu(x) \sim \frac{1}{2} \Gamma(\nu) \left ( \frac{1}{2} \; x \right 
)^{-\nu} \; \;, \; \; \; {\rm for} \; \; x\sim 0
\end{equation}

\noindent
which implies that $K_\nu(x)$ has a singularity at $x=0$. This reveals that
$P(x)$ is singular at $x=0$. Our strategy here is to remove this singularity
by having an appropriate relation between the two remaining constants $B$ and
$D$ in the solution (36). To derive this relation we proceed as follows:

We first derive the asymptotic expansion for the solution of the modified 
Bessel equation (30) whose leading behavior is (31a). To do this we peel off
the leading behavior by substituting
\begin{equation}
y(x) = {\cal C}_1 \; x^{-1/2} \; e^x \; w(x)
\end{equation}

\noindent
into the modified Bessel equation (30). The equation satisfied by $w(x)$ is 
\begin{equation}
x^2\; w''(x) + 2x^2\; w'(x) + \left ( \frac{1}{4} -\nu^2\right 
)\; w(x) = 0 \; \; .
\end{equation}

\noindent
We seek a solution of this equation of the form $w(x)=1+\varepsilon(x)$ with
$\varepsilon (x) \ll 1$ ($x\rightarrow\infty$). $\varepsilon(x)$ satisfies the 
equation 
\begin{equation}
x^2\; \varepsilon''(x) + 2x^2\; \varepsilon'(x) + 
\left ( \frac{1}{4} -\nu^2\right 
)\; \varepsilon (x) + \left ( \frac{1}{4} -\nu^2\right ) = 0 \; \; ,
\end{equation}

\noindent
which may be simplified by the approximations
\begin{equation}
\left ( \frac{1}{4} - \nu^2 \right ) \varepsilon \; \ll \; \frac{1}{4} - \nu^2
\; \; , \; \;  x^2\varepsilon'' \; \ll \; x^2 \varepsilon' \; \; ; \; \;
x\rightarrow \infty \; \; .
\end{equation}

\noindent
We make the second of these approximations because we anticipate that
$\varepsilon$ decays like a power of $x$ as $x\rightarrow\infty$. The 
resulting asymptotic differential equation is
\begin{eqnarray*}
2x^2\; \varepsilon' \sim \left ( \nu^2 - \frac{1}{4} \right ) 
\; \; , \; \; x\rightarrow\infty \; \; .
\end{eqnarray*}

Ordinarily the solution to this equation would be 
$\varepsilon(x) \sim {\tilde{c}}$ as $x\rightarrow\infty$ where ${\tilde{c}}$ 
is an integration constant. However since
$\varepsilon(x) \ll 1$ as $x\rightarrow\infty$ we must set ${\tilde{c}}=0$. 
The leading behavior of $\varepsilon(x)$ is then given by
\begin{eqnarray*}
\varepsilon(x) \sim \left ( \frac{1}{8} -\frac{1}{2} \nu^2 \right )\; x^{-1}
\; \; , \; \; x\rightarrow\infty \; \; .
\end{eqnarray*}

\noindent
This kind of analysis \cite{bo} can be repeated to obtain all the terms in the 
asymptotic expansion of $w(x)$ as $x\rightarrow\infty$. However the leading
behavior of $\varepsilon(x)$ suggests that $w(x)$ has a series expansion in
inverse power of $x$. Thus to simplify the analysis we assume at the outset
that
\begin{equation}
w(x) \sim \sum_{n=0}^\infty a_n \; x^{-n} \; \; , \; \; (x\rightarrow\infty \;
, \; a_0=1) \; \; .
\end{equation}

\noindent
Substituting this expression into the differential equation for $w(x)$ gives
\begin{equation}
\sum_{n=0}^\infty n(n+1) \; a_n \; x^{-n} - 2\sum_{n=0}^\infty n\; a_n\;
x^{1-n} + \left ( \frac{1}{4} -\nu^2 \right ) \sum_{n=0}^\infty
a_n \; x^{-n} \sim 0 \; \; .
\end{equation}

\noindent
Since the coefficients of any asymptotic power series are unique we equate
to zero the coefficients of all power of $\frac{1}{x}$ in the above relation
\begin{equation}
x^{-n} \; : \; \left [ \left ( n+\frac{1}{2} \right )^2 -\nu^2 \right ] a_n
- 2 \; (n+1) \; a_{n+1} =0 \; \; , \; \; n=0, 1, 2 \ldots \; \; .
\end{equation}

\noindent
Solving this recursion relation and using $a_0=1$ we obtain
\begin{equation}
w(x) \sim 1 - \frac{(4\nu^2-1^2)}{1! \; 8x} + 
\frac{(4\nu^2-1^2)(4\nu^2-3^2)}{2! \; (8x)^2} - \ldots \; \; , \; \;
x\rightarrow\infty \; \; .
\end{equation}

\noindent
From the ratio test we see that the radius of convergence $R$ of (41) is
\begin{equation}
R = \lim_{n\rightarrow\infty} \; \left |\frac{a_n}{a_{n+1}}\right | =
\lim_{n\rightarrow\infty} \; \frac{2 (n+1)}{(n+1)^2-\nu^2} = 0
\end{equation}

\noindent
unless the series (41) terminates, which it does when
\begin{eqnarray*}
\nu = \pm \frac{1}{2} \; , \; \pm \frac{3}{2} \; , \; \pm \frac{5}{2} \;
, \; \ldots \; \; .
\end{eqnarray*}

\noindent
When this happens, the finite series (41) when multiplied by $e^{-x}/\sqrt{x}$
gives an {\it exact} solution to the modified Bessel equation.

Similarly, the complete asymptotic series for the function whose leading
behavior is given by (31b) is
\begin{eqnarray*}
y(x) \sim {\cal C}_2 \; x^{-1/2} \; e^{-x} \; w(x)
\end{eqnarray*}
\noindent
where
\begin{equation}
w(x) \sim 1 + \frac{(4\nu^2-1^2)}{1! \; 8x} + 
\frac{(4\nu^2-1^2)(4\nu^2-3^2)}{2! \; (8x)^2} + \ldots \; \; , \; \;
x\rightarrow\infty \; \; .
\end{equation}

\noindent
By global analysis \cite{bo} the two constants ${\cal C}_1$ and ${\cal C}_2$ 
can be derived as
\begin{eqnarray}
{\cal C}_1 & = &  (2\pi)^{-1/2} \nonumber \\
{\cal C}_2 & = & \sqrt{\pi/2} \; \; ,
\end{eqnarray}

\noindent
to get the modified Bessel functions $I_\nu(x)$ and $K_\nu(x)$ respectively.

Thus, in general, $I_\nu(x)$ and $K_\nu(x)$ are represented by infinite
power series. The above analysis reveals that for 
$\nu = \pm 1/2, \pm3/2, \ldots$, the series terminates and from the 
asymptotic expansion we get an {\it exact} solution. In such a case, i.e., 
when the power series terminates it is possible to find a relation between
the two constants $B$ and $D$ of the general solution (\ref{gsol}), such that the
singularity of $P(x)$ at $x=0$ is removed. {\it Thus both the boundary
conditions (i) and (ii)} [i.e., $P(x)=0$ for $|x|\rightarrow\infty$ and
$P(x)$ is finite at $x=0$] {\it are satisfied by the solution (\ref{gsol}) provided
$\nu$ is an odd half integers, i.e.,}
\begin{equation}
\nu = n+\frac{1}{2} \; \; , \; \; n=0,1,2,\ldots \; \; .
\end{equation}

\noindent
By Eq.(25) we have
\begin{eqnarray*}
\frac{1}{2} \left [ 1+ \frac{\beta c_{01}}{\omega_0^2 c_2} \right ] = 
n + \frac{1}{2} \; \; , \; \; n=0,1,2,\ldots \; \; .
\end{eqnarray*}

\noindent
But $n=0$ implies $c_{01} =0$ or $c_0=c_1$ which is not allowed physically.
Therefore we have
\begin{equation}
\label{fdr}
\frac{\beta \; c_{01} }{2\omega_0^2 \; c_2}  = n \; \; ; \; \; 
n=1,2,3,\ldots \; \; ,
\end{equation}

\noindent
In passing, we mention here that the integers $n$ characterize the distinct 
physically allowed steady states of the thermodynamically open systems.

With the above mentioned restricted values of $\nu$ 
($\nu = n+\frac{1}{2} \; , \; n=1,2,3,\ldots$) we now explicitly calculate
the several quantities [ details are given in Appendix-B ]which appeared in 
the general solution $P(x)$ in Eq.(\ref{gsol}). We finally obtain :
\begin{eqnarray}
P_{n+\frac{1}{2}}(x) & = & B \; \sqrt{\frac{\pi}{2\;b^{1/2}}} 
\sum_{k=0}^n f_k^n \;\frac{e^{-\sqrt{b}x}}{x^{k+n+1}} \nonumber \\
& & -\frac{D}{2\sqrt{b}} \; \sum_{i=0}^n \sum_{k=0}^n \sum_{j=0}^{n-k} 
\left \{ (-1)^i + (-1)^{j-n} \right \} \; f_i^n \; f_k^n \; 
\frac{(n-k)!}{j!} \; \frac{x^{j-i-n-1}}{(\sqrt{b})^{n-k-j+1}} \; \; 
\end{eqnarray}

\noindent
where we have defined
\begin{equation}
f_k^n = \frac{ (n+k)! }{ 2^k \; b^{k/2} \; k! \; (n-k)! } \; \; .
\end{equation}

We are now in a position to remove the singularity of $P(x)$ at $x=0$ . To
achieve this we seek a relation between the constants $B$ and $D$ . For
this we now expand the exponential in Eq.(47) . It is convenient to write
a few nontrivial steps explicitly after this expansion. We thus have
\begin{eqnarray}
P_{n+\frac{1}{2}}(x) & = & B \; \sqrt{\frac{\pi}{2\;b^{1/2}}} 
\sum_{k=0}^n \sum_{\alpha =0}^\infty (-1)^\alpha \; 
\frac{(\sqrt{b})^\alpha} {\alpha !} \;
\frac{f_k^n }{x^{k+n+1-\alpha}} \nonumber \\
& & -\frac{D}{2\sqrt{b}} \; \sum_{i=0}^n \sum_{k=0}^n \sum_{j=0}^{n-k} 
(-1)^i \; f_i^n \; f_k^n \; 
\frac{(n-k)!}{j!} \; \frac{1}{(\sqrt{b})^{n-k-j+1}} 
\; \frac{1}{x^{i+n+1-j}}  \nonumber \\
& & -\frac{D}{2\sqrt{b}} \; \sum_{i=0}^n \sum_{k=0}^n \sum_{j=0}^{n-k} 
(-1)^{j-n} \; f_i^n \; f_k^n \; 
\frac{(n-k)!}{j!} \; \frac{1}{(\sqrt{b})^{n-k-j+1}} 
\; \frac{1}{x^{i+n+1-j}}  \nonumber \\
& = &  B \; \sqrt{\frac{\pi}{2\;b^{1/2}}} 
\sum_{k=0}^n \sum_{\alpha =0}^\infty (-1)^\alpha \; 
\frac{(\sqrt{b})^\alpha} {\alpha !} \;
\frac{f_k^n }{x^{k+n+1-\alpha}} \nonumber \\
& & -\frac{D}{2\sqrt{b}} \; \sum_{i=0}^n  \sum_{j=0}^n 
(-1)^i \; f_i^n \; f_0^n \; 
\frac{n!}{j!} \; \frac{1}{(\sqrt{b})^{n-j+1}} 
\; \frac{1}{x^{i+n+1-j}}  \nonumber \\
& & -\frac{D}{2\sqrt{b}} \; \sum_{i=0}^n \sum_{j=0}^{n-1} 
(-1)^i \; f_i^n \; f_1^n \; 
\frac{(n-1)!}{j!} \; \frac{1}{(\sqrt{b})^{n-1-j+1}} 
\; \frac{1}{x^{i+n+1-j}}  \nonumber \\
& & -\frac{D}{2\sqrt{b}} \; \sum_{i=0}^n \sum_{j=0}^{n-2} 
(-1)^i \; f_i^n \; f_2^n \; 
\frac{(n-2)!}{j!} \; \frac{1}{(\sqrt{b})^{n-2-j+1}} 
\; \frac{1}{x^{i+n+1-j}}  \nonumber \\
& & - \; \ldots  \nonumber \\
& & -\frac{D}{2\sqrt{b}} \; \sum_{i=0}^n \sum_{j=0}^n 
(-1)^{j-n} \; f_i^n \; f_0^n \; 
\frac{n!}{j!} \; \frac{1}{(\sqrt{b})^{n-j+1}} 
\; \frac{1}{x^{i+n+1-j}} \nonumber \\
& & -\frac{D}{2\sqrt{b}} \; \sum_{i=0}^n \sum_{j=0}^{n-1} 
(-1)^{j-n} \; f_i^n \; f_1^n \; 
\frac{(n-1)!}{j!} \; \frac{1}{(\sqrt{b})^{n-1-j+1}} 
\; \frac{1}{x^{i+n+1-j}}  \nonumber \\
& & - \; \ldots  \nonumber \\
& & ({\rm where \; we \; have \; carried \; out \; the \; summation \; over}
\; k) \nonumber \\
& = & B \; \sqrt{\frac{\pi}{2\;b^{1/2}}} 
\sum_{k=0}^n \sum_{\alpha =0}^\infty (-1)^\alpha \; 
\frac{(\sqrt{b})^\alpha} {\alpha !} \;
\frac{f_k^n }{x^{k+n+1-\alpha}} \nonumber \\
& & -\frac{D}{2\sqrt{b}} \; \sum_{k=0}^n \left [ 
\sum_{j=0}^n  (-1)^k \; f_k^n \; f_0^n \; \frac{n!}{j!} \; 
\frac{1}{(\sqrt{b})^{n-j+1}} 
\; \frac{1}{x^{k+n+1-j}} \right.  \nonumber \\
& & \left. + \; \sum_{j=0}^{n-1}  (-1)^k \; f_k^n \; f_1^n \; 
\frac{(n-1)!}{j!} \; \frac{1}{(\sqrt{b})^{n-j}} 
\; \frac{1}{x^{k+n+1-j}} + \; \ldots \right ]   \nonumber \\
& & -\frac{D}{2\sqrt{b}} \; \sum_{k=0}^n \left [ 
\sum_{j=0}^n  (-1)^{j-n} \; f_k^n \; f_0^n \; \frac{n!}{j!} \; 
\frac{1}{(\sqrt{b})^{n-j+1}} 
\; \frac{1}{x^{k+n+1-j}} \right.  \nonumber \\
& & \left. + \; \sum_{j=0}^{n-1}  (-1)^{j-n} \; f_k^n \; f_1^n \; 
\frac{(n-1)!}{j!} \; \frac{1}{(\sqrt{b})^{n-j}} 
\; \frac{1}{x^{k+n+1-j}}  \; + \; \ldots \right ]
\end{eqnarray}

\noindent
(where we have replaced the dummy index $i$ by $k$ in the $D$-containing 
sums)

To remove the singularity at $x=0$ from the above expression for 
$P_{n+\frac{1}{2}}(x)$, $B$ must be related to $D$ in such a way that all 
powers of $\frac{1}{x}$ vanishes identically. For this the coefficients of
$\frac{1}{x^{k+1}}$ from the summation are equated to zero. This gives
\begin{equation}
B\; \sqrt{\frac{\pi}{2\;b^{1/2}}}\; (-1)^n \; \frac{(\sqrt{b})^n}{n!} \; 
f_k^n - \frac{D}{2\sqrt{b}} \; f_k^n \; f_0^n \; \frac{1}{\sqrt{b}}
- \frac{D}{2\sqrt{b}} \; f_k^n \; f_0^n \; \frac{1}{\sqrt{b}} = 0 \; \; .
\end{equation}

\noindent
Note that the relevant coefficients contribute from the $B$-containing term
for $\alpha=1$ and from $D$-containing term for $j=n$. Putting the value
of $f_0^n$ from Eq.(48) in Eq.(50) we obtain the desired relation between $B$
and $D$.
\begin{equation}
B = (-1)^n \; D \; \frac{n!}{\sqrt{\pi}} \; 2^{(2n+1)/2} \;
\frac{1}{b^{(2n+3)/4}} \; \; .
\end{equation}

From Eq.(51) it reveals that the relation between $B$ and $D$ is independent
of any dummy index $i$, $j$ or $k$ and only depends on $n$. Thus for any $n$ 
the relation is unique. With this choice the probability 
$P_{n+\frac{1}{2}}(x)$ (Eq.(47)) becomes finite for all $x$ and is given by
\begin{eqnarray}
P_{n+\frac{1}{2}}(x) & = & (-1)^n \; D \; 2^n \; n! \; \frac{1}{b^{(n+2)/2}}
\sum_{k=0}^n f_k^n \; e^{-\sqrt{b} x} \frac{1}{x^{k+n+1}} \nonumber \\
& & - \; \frac{D}{2\sqrt{b}} \sum_{i=0}^n \sum_{k=0}^n \sum_{j=0}^{n-k}
\{ (-1)^i + (-1)^{j-n} \} \; f_i^n \; f_k^n \; \frac{(n-k)!}{j!} \;
\frac{x^{j-i-n-1}}{(\sqrt{b})^{n-k-j+1}}
\end{eqnarray}

\noindent
The above probability distribution function which remains undetermined upto
the constant $D$ (which we are to determine shortly) vanishes as 
$x\rightarrow\infty$ and remains finite at the barrier top $x=0$. For 
illustration, we explicitly write some of the probability distribution 
functions. 

\noindent
For $n=1$ case we have,
\begin{mathletters}
\begin{eqnarray}
P_{3/2}(x) & = & -2D \; b^{-3/2} \; e^{-\sqrt{b} x} \; \left ( \frac{1}{x^2}
+\frac{1}{\sqrt{b} x^3} \right ) - \frac{D}{b} \; \left ( \frac{1}{x} -
\frac{2}{bx^3} \right ) \; \; ; \; \; x\; >\; 0 \\
& = &  -2D \; b^{-3/2} \; \; e^{\sqrt{b} x} \; \left ( \frac{1}{x^2}
- \frac{1}{\sqrt{b} x^3} \right ) + \frac{D}{b} \; \left ( \frac{1}{x} -
\frac{2}{bx^3} \right ) \; \; ; \; \; x\; <\; 0  \\
& = & -\frac{2D}{3\sqrt{b}} \; \; \; \; ; \; \; \; \; x=0
\end{eqnarray}
\end{mathletters}

\noindent
where the probability distributions for $x<0$ are obtained by noting the 
symmetry of the differential equation.


To normalize the probability and thereby to calculate the current over the 
top of the barrier located at $x=0$, we consider the condition for 
normalization:
\begin{equation}
\int_{-\infty}^{+\infty} P_{n+\frac{1}{2}} (x) \; dx = 1 \; \; .
\end{equation}

\noindent
Since the probability decreases rapidly (due to the presence of the modified  
Bessel function $K_\nu (x)$) we approximate the integral (54) by
\begin{equation}
\int_{-\Delta}^{+\Delta} P_{n+\frac{1}{2}} (x) \; dx = 1 \; \; ; \; \;
\Delta \; {\rm large \; but \; finite} 
\end{equation}

\noindent
where $\pm \Delta$ approximately indicates the two zeroes of the inverted 
potential $V(x)$. By Eq.(17), $\Delta$ may be expressed in terms of $E_0$ 
the height of the barrier as,
\begin{eqnarray*}
\Delta = (2E_0 / \omega_0^2)^{1/2} \; \; .
\end{eqnarray*}

\noindent
By symmetry we rewrite
\begin{equation}
\int_0^{\Delta} P_{n+\frac{1}{2}} (x) \; dx = 1/2 \; \; ; \; \;
\Delta \; {\rm large \; but \; finite}  \; \; .
\end{equation}

Due to the presence of the powers of $\frac{1}{x}$ in the expression for 
$P_{n+\frac{1}{2}}(x)$, it is not possible to carry out the integration in
Eq.(56) in a straightforward manner. We therefore employ a limiting procedure
and make use of the expression (52) for $P_{n+\frac{1}{2}}(x)$ in (56) to 
obtain an expression for the following normalization constant $D$
\begin{equation}
D = \frac{b}{
2 \; {\rm Ein}(\sqrt{b} \Delta) + 2^{n+1} \; n! \; \sum_{k=0}^n 
\sum_{j=0}^{k+n-1} \frac{(-1)^k}{2^k \; k! \; (n-k)! \; (n+k-j)}
} \; \; ,
\end{equation}

\noindent
where $\Delta$ is large but finite and the function Ein$(x)$ is defined as
\cite{as}
\begin{equation}
{\rm Ein}(x) = \sum_{k=1}^\infty (-1)^k \; \frac{x^k}{k! \; k} \; \; .
\end{equation}

The expression (52) together with the normalization constant $D$ as given by
(57) yields the complete and exact analytical expression for the 
probability distribution function $P_{n+\frac{1}{2}}(x)$. In 
Fig.(\ref{fig2}) we draw
two typical normalized probabilities for $n=1$ and $n=2$ for the 
parametric values $\beta=1.0$, $\Delta=2.5$ and $c_2=0.9$. The effect of
third order noise is illustrated in Fig.(\ref{fig3}), where we exhibit the
probability distribution functions for several values of the third order
noise strength $c_2$. One observes that the distribution gets flattened as
$c_2$ increases as expected. 


\section{ Dynamics of barrier crossing induced by adiabatic noise }

We now extend the above analysis to elucidate the following problem of
dynamics of barrier crossing.

An overdamped particle moves in an external field of force and in addition to 
this is subject to an adiabatically fluctuating force of external origin. The
conditions are such that the particle is originally caught in the potential
well in $V(x)$,
but may escape in course of time over the barrier. Our object here is to 
calculate the rate of escape from this well. The analogous problem for the
case of fast fluctuations is the celebrated Kramers' problem of Brownian 
motion in phase space \cite{kramers}. Our calculation rests on the third order equation of
motion (9) obeyed by probability distribution function derived in Sec.~{II}.

To proceed further we employ the
The popular flux-over-population method originated by Farkas \cite{farkas}
many years ago.
The calculation rests on the evaluation of two quantities;
(i) the steady state current $J$ over the barrier top, (located at $x=0$) 
that results if the particles are continuously fed into the domain
of attraction (say, in the region of left well) and are subsequently and 
continuously removed in the neighboring domain of attraction. (ii) Steady 
state population $n_a$ in the initial domain of attraction, i.e., the 
left well. The rate is defined by
\begin{equation}
{\cal K} = J/n_a \; \; .
\end{equation}

\noindent
The method has been used by Kramers in his seminal work on barrier crossing 
dynamics and many others over the several decades 
\cite{kramers,rate,melnikov}.

\subsection{Calculation of $n_a$}

For calculation of $n_a$ it is necessary to evaluate the stationary 
probability density $P_b(x)$ near the bottom of the well
corresponding to a zero current ($J=0$) situation along
the $x$ co-ordinate. 

We first linearize the potential $V(x)$ as shown in Fig.(\ref{fig1}) near the
bottom of the left well, at $x=-\Delta$, so that we approximate
\begin{equation}
V(x) \simeq \frac{1}{2} \; \omega_b^2 \; (x+\Delta)^2 \; \; ,
\end{equation}

\noindent
where $\omega_b$ refers to the frequency at the bottom of the left well. 
Eq.(60) and $J=0$ condition reduce the third order equation of motion (Eq.(9))
for probability density $P_b(x)$ inside the well to the following form:
\begin{equation}
(x+\Delta) \; \frac{d^2 P_b}{dx^2} + \gamma' \; \frac{d P_b}{dx} -
b \; (x+\Delta) \; P_b = 0 \; \; .
\end{equation}

\noindent
The above equation is valid near the bottom of the left well 
($x \simeq -\Delta$). Here $\gamma'$ is defined as
\begin{equation}
\gamma' = \frac{2\omega_b^2 - a}{\omega_b^2} \; \; ,
\end{equation}

\noindent
where $a$ is as given by Eq.(16). Putting $z=x+\Delta$ and $y=zP_b$, 
Eq.(61) can be transformed as follows :
\begin{equation}
z^2 \; \frac{d^2 y}{dz^2} -(2- \gamma') \; z \; \frac{d y}{dz} +
[(2-\gamma') - b \; z^2] \; y = 0 \; \; .
\end{equation}

\noindent
From Eqs.(62) and (16) we have
\begin{equation}
2-\gamma' = \frac{\beta \; c_{01}}{c_2 \; \omega_b^2} \; \; .
\end{equation}

\noindent
Note that $2-\gamma'$ is a positive quantity. We write
\begin{equation}
2-\gamma' = \sigma \; \; .
\end{equation}

\noindent
Therefore Eq.(63) is given by
\begin{equation}
z^2 \; \frac{d^2 y}{dz^2} - \sigma \; z \; \frac{d y}{dz} +
[\sigma - b \; z^2] \; y = 0 \; \; .
\end{equation}

\noindent
Substitution of $y(z) = z^{\frac{1}{2} (\sigma +1)} \; W(z)$ in Eq.(66) yields
\begin{equation}
z^2 \; \frac{d^2 W}{dz^2} + z \; \frac{d W}{dz} -
[\nu^{\prime 2}+ b \; z^2] \; W = 0 \; \; ,
\end{equation}

\noindent
where
\begin{equation}
\nu' = \frac{\sigma-1}{2} \; \; .
\end{equation}

The solutions of Eq.(67) are again the modified Bessel functions, 
$I_{\nu'} (\sqrt{b} z)$ and $K_{\nu'} (\sqrt{b} z)$. Reverting back to 
original variables, the general solution for the steady state probability
distribution near the bottom of the left well is given by
\begin{equation}
P_b(x) = A' \; (x+\Delta)^{\nu'} \; I_{\nu'} [\sqrt{b}\; (x+\Delta)] +
B' \; (x+\Delta)^{\nu'} \; K_{\nu'} [\sqrt{b}\; (x+\Delta)] \; \; ,
\end{equation}

\noindent
$A'$ and $B'$ are the two arbitrary constants of integration. 

By demanding that $P_b(x)$ must vanish at infinity, we require
\begin{equation}
A' = 0
\end{equation}

\noindent
and therefore
\begin{equation}
P_b(x) = B' \; (x+\Delta)^{\nu'} \; K_{\nu'} [\sqrt{b}\; (x+\Delta)] \; \; .
\end{equation}

\noindent
Furthermore, we note that although $K_{\nu'}$ itself is singular at 
$x=-\Delta$ the presence of $(x+\Delta)^{\nu'}$ in Eq.(71) assures that
the probability (71) remains finite at $x=-\Delta$. To verify this
assertion, we use the property of modified Bessel function $K_{\nu'}(z)$
for fixed $\nu'$ and $z\rightarrow 0$. $K_{\nu'}(z)$ behaves as \cite{as}
\begin{equation}
K_{\nu'}(z) \sim \frac{1}{2} \; \Gamma(\nu') \; \left ( \frac{1}{2}
z \right )^{-\nu'} \; \; , \; \; {\rm Re} \; \nu' \; > \; 0 \; \; .
\end{equation}

\noindent
Hence
\begin{eqnarray}
P_b(-\Delta) & = & B' \; \lim_{x\rightarrow -\Delta} \; (x+\Delta)^{\nu'} \;
K_{\nu'} [\sqrt{b} (x+\Delta)] \nonumber\\
& = & B' \; \frac{\Gamma(\nu') \; 2^{\nu' -1}}{b^{\nu'/2}} \; \; .
\end{eqnarray}

\noindent
We thus see that the steady state probability $P_b$ 
( given by Eq.(71) ) is finite at the bottom of the left well.

\noindent
The above solution $P_b(x)$ must now be subject to the following boundary 
condition :
\begin{equation}
P_b(-\Delta) = P_t^{n+\frac{1}{2}} (-\Delta)
\end{equation}

\noindent
where the stationary probability $P_t^{n+\frac{1}{2}} (-\Delta)$ corresponds
to the vanishing current $J=0$ along $x$ pertaining to the homogenous version
of Eq.(20). As usual, $P_t^{n+\frac{1}{2}} (x)$  must also satisfy the
boundary condition that for $|x| \rightarrow \infty$, 
$P_t^{n+\frac{1}{2}} (x)$ vanishes. Such a solution is immediately apparent
from our earlier analysis of Sec.~{III}. Thus
\begin{mathletters}
\begin{eqnarray}
P_t^{n+\frac{1}{2}} (x) & = & B\; \sqrt{\frac{\pi}{2\; b^{1/2}}} \;
\sum_{k=0}^n f_k^n \; \frac{e^{-\sqrt{b} x}}{x^{k+n+1}} \; \; ; \; \;
x \; > \; 0 \\
& = & B\; \sqrt{\frac{\pi}{2\; b^{1/2}}} \;
\sum_{k=0}^n (-1)^{k+n+1} \; f_k^n \; \frac{e^{\sqrt{b}x}}{x^{k+n+1}} 
\; \; ; \; \; x \; < \; 0 
\end{eqnarray}
\end{mathletters}

\noindent
where $f_k^n$ is as defined in Eq.(48).

\noindent
Making use of Eqs.(73) and (75) in Eq.(74) we obtain a relation between $B$
and $B'$.
\begin{equation}
B' = B \; \sqrt{\frac{\pi}{2\; b^{1/2}}} \; 
\frac{b^{\nu'/2}}{\Gamma(\nu') \; 2^{\nu' -1}} \; \sum_{k=0}^n 
(-1)^{k+n+1} \; f_k^n \; \frac{e^{\sqrt{b} \Delta}}{\Delta^{k+n+1}} \; \; .
\end{equation}

\noindent
Therefore $P_b(x)$ in Eq.(71) may be expressed as
\begin{equation}
P_b(x) = B \; \sqrt{\frac{\pi}{2\; b^{1/2}}} \; 
\frac{b^{\nu'/2}}{\Gamma(\nu') \; 2^{\nu' -1}} \; \sum_{k=0}^n 
(-1)^{k+n+1} \; f_k^n \; \frac{e^{\sqrt{b} \Delta}}{\Delta^{k+n+1}} 
(x+\Delta)^{\nu'} \; K_{\nu'} [ \sqrt{b} (x+\Delta) ] \; .
\end{equation}

\noindent
The above distribution which is valid near the bottom of the left well may be 
used to calculate the population inside the left well as,
\begin{equation}
n_a = 2 \int_{-\Delta}^0 P_b(x) \; dx \; \; .
\end{equation}

\noindent
Due to the presence of $K_{\nu} (x)$ the probability $P_b(x)$ is a 
rapidly decreasing function. We may extend the above integration limit to
infinity. This yields ( using Eq.(71) ) \cite{gr}
\begin{equation}
n_a = B' \; \frac{ \sqrt{\pi}\; 2^{\nu'} \; \Gamma(\nu' +\frac{1}{2}) }
{ (\sqrt{b} )^{\nu' +1} } \; \; .
\end{equation}

\noindent
Using the relations (76) and (79) we finally have
\begin{equation}
n_a = \sqrt{2}\; \pi \; B \; \frac{ \Gamma (\nu' +\frac{1}{2})}{\Gamma (\nu')} \;
\sum_{k=0}^n (-1)^{k+n+1} \; f_k^n \; 
\frac{  e^{\sqrt{b} \Delta}}{\Delta^{k+n+1}} \; b^{-3/4} \; \; .
\end{equation}

\subsection{Calculation of escape rate}

Having determined the population $n_a$ of the left well and the steady
state current $J$  from Eqs.(80) and (19), respectively we are now in a
position to calculate the escape rate in terms of the Eq.(59). We thus obtain
(for calculation of $J$ the linearization of the potential $V(x)$ at
$x=0$ which results in Eq.(17) has been carried out)
\begin{eqnarray}
{\cal K}_{n+\frac{1}{2}} & =  & J/n_a  \nonumber \\
& = &  - \frac{c_2 \; \omega_0^2}{\sqrt{2}\; \pi \; \beta^3} \; \left (
\frac{D}{B} \right ) \; \frac{\Gamma(\nu')}{\Gamma(\nu' +\frac{1}{2})} \;
\left \{ \frac{b^{3/4}}{
\sum_{k=0}^n (-1)^{k+n+1} \; f_k^n \; 
\frac{  e^{\sqrt{b} \Delta}}{\Delta^{k+n+1}} \; \; } \right \} \; \; .
\end{eqnarray}

\noindent
Making use of the relation (51) between the two constants $B$ and $D$ in
Eq.(81) one finds
\begin{equation}
{\cal K}_{n+\frac{1}{2}}  =  (-1)^{n+1} \; 
\frac{c_2 \; \omega_0^2}{\sqrt{2\pi} \; \beta^3} \; 
\frac{\Gamma(\nu')}{\Gamma(\nu' +\frac{1}{2})} \; \frac{1}{n!} \; 
\frac{1}{2^{n+\frac{1}{2}}} \;
\left \{ \frac{b^{(n+3)/2}}{
\sum_{k=0}^n (-1)^{k+n+1} \; f_k^n \; 
\frac{  e^{\sqrt{b} \Delta}}{\Delta^{k+n+1}} \; \; } \right \} \; \; .
\end{equation}

\noindent
The expression (82) can be simplified further by noting the following
relations. First, we have from Eqs.(19) and (23) 
\begin{mathletters}
\begin{equation}
\nu = \frac{1}{2} \; (\gamma -1) \; \; \; \; {\rm where} \; \;
\gamma = 2+\frac{a}{\omega_0^2} \; \; ,
\end{equation}

\noindent
and from Eqs.(62), (65) and (68)
\begin{equation}
\nu' = \frac{1}{2} \; (1- \gamma') \; \; \; \; {\rm where} \; \;
\gamma' = 2 - \frac{a}{\omega_b^2} \; \; .
\end{equation}
\end{mathletters}

\noindent
Thus we have
\begin{equation}
\nu + \nu' = \frac{1}{2} \; a \; \left ( 
\frac{\omega_0^2+\omega_b^2}{\omega_0^2 \; \omega_b^2} \right ) \; \; .
\end{equation}

\noindent
Furthermore Eqs.(16) and Eq.(\ref{fdr}) suggest that 
\begin{equation}
a = 2 \; n \; \omega_0^2 \; \; ; \; \; b = \frac{\beta^2}{c_2} \; \; ;
\; \; n=1,2,\ldots
\end{equation}

\noindent
Use of Eq.(85) in Eq.(83b) yields
\begin{equation}
\nu' = n\; \frac{\omega_0^2}{\omega_b^2} - \frac{1}{2} \; \; ; \; \;
n=1,2,\ldots
\end{equation}

\noindent
The above relation together with the expression for $f_k^n$ (Eq.(48))
leads to the
formula for the transition rate ${\cal K}_{n+\frac{1}{2}}$ as follows :
\begin{equation}
{\cal K}_{n+\frac{1}{2}}  = 
\frac{c_2 \; \omega_0^2}{\sqrt{\pi} \; \beta^3} \; \frac{1}{n!} \; 
\frac{\Gamma(n\; \frac{\omega_0^2}{\omega_b^2} - \frac{1}{2} )}
{\Gamma(n\; \frac{\omega_0^2}{\omega_b^2} )} \;  
\left \{ \;
\frac{e^{-\sqrt{b} \Delta} }{
\sum_{k=0}^n (-1)^k \; \frac{(n+k)!}{k! \; (n-k)!} \; 
\frac{2^{n-k+1}}{\Delta^{k+n+1}} \; \frac{1}{b^{(n+k+3)/2}} 
} \; 
\right \} \; \; .
\end{equation}

\noindent
$\Delta$-s refer to the zero's of the potential $V(x)$ as shown 
in Fig.(\ref{fig1}) and by
virtue of linearization of $V(x)$ at $x=0$ (Eq.(17)) $\Delta$ is
approximately given by
\begin{equation}
\Delta = \left ( \frac{2\; E_0}{\omega_0^2} \right )^{1/2} \; \; .
\end{equation}

The above expression can be made more transparent by demonstrating a 
representative transition rate, say, for $n=1$
as follows :
\begin{equation}
{\cal K}_{3/2} = \frac{1}{8\sqrt{\pi}} \; \frac{c_{01}}{c_2^2} \;
\frac{\Gamma (\frac{\beta \; c_{01}}{2\; c_2 \; \omega_b^2} -\frac{1}{2} )}
{\Gamma (\frac{\beta \; c_{01}}{2\; c_2 \; \omega_b^2} )} \;
\Delta^2 \; \beta^2 \exp \left (-\frac{\beta\; \Delta}{\sqrt{c_2}} \right )
\end{equation}


The above expressions are analogous to Kramers' formula for the rate of escape 
from a potential well over a finite barrier of height $E_0$ under the
influence of an external nonthermal adiabatic noise. What we have shown here
is that within a linearized description of the potential, the corresponding 
diffusion process can be described exactly for dissipation $\beta$ pertaining 
to the timescale $\frac{1}{\beta}\; \ll \; \Delta t\; \ll \;\tau_c$ . 

The escape rate expressions derived above suggest that the rate approaches 
zero both for $\beta \rightarrow \infty$ and $\beta \rightarrow 0$. This 
behavior is somewhat reminiscent of Kramers' theory, where it was noted 
earlier that these two limiting behaviors imply a maximal rate at some
damping value $\beta$. The rate therefore undergoes a turnover in a form of a
bell-shaped curve. In Fig.(\ref{fig4}) we plot a representative variation of the
escape rate versus dissipation $\beta$ for different third order noise 
strength. With increasing friction, the rate undergoes a turnover from an
increasing behavior at low friction to an inverse behavior in the high 
friction limit. 

Since the driving noise is of {\it nonthermal} origin, the escape rate 
expression is devoid of any temperature. Temperature is characteristic of 
a closed thermodynamic system at equilibrium. What we have  here instead is 
the ratio of strength of nonthermal noise $\sqrt{c_2}$ to dissipation $\beta$ 
in the exponential factor of the rate expressions. As Ma \cite{ma} pointed 
out that in the steady state such a parameter might play the role of 
temperature in the open system which characterizes the steady state.

As emphasized earlier that since the noise is {\it external}
and the noise and dissipation have no common mechanistic origin 
(in contrast to what one observes in theory of Brownian motion) 
the steady state is not allowed arbitrarily. The system can attain the steady 
states depending on the specific integers $n$ ($n=1,2,\ldots$) which uniquely 
connect the parameters $c_0$, $c_1$, $c_2$ according to the 
relation derived in Sec.~{III}.


\section{Conclusions}

In this paper we have presented a solution for the problem of motion of a 
particle in a force field simultaneously subject to an external adiabatic 
noise characterized by long correlation time without keeping any restriction 
on the type of noise that it is Gaussian. Specifically we have calculated 
the rate of escape of the particle over the barrier initially confined in a 
well induced by nonthermal fluctuations. The theory rests on a perturbative 
expansion in $|\mu|^{-1}$ ( where $|\mu|$ is the damping constant or the 
largest eigenvalue of the unperturbed system ) pertaining to the separation 
of timescales (\ref{ineqb}) as carried out in our earlier papers 
\cite{jpa1,jpa2}. The main conclusions of our study are as follows:

(i) An analogue of Smoluchowski equation in the case of adiabatic 
non-Gaussian noise processes (Eq.(9)) has been proposed.

(ii) The undergoing stochastic process is characterized by third order noise
which is responsible for non-Gaussian features \cite{jpa2}.

(iii) Given the appropriate boundary conditions the third order equation
admits of exact solution for the linearized potential.

(iv) The interplay of the characteristic linear dissipation of the system
and the external noise leads to physically allowed distinct steady states 
subject to appropriate boundary conditions. 

(v) In the spirit of Kramers theory we have solved the problem of escape
of the particle confined in a well and have shown that the escape
rate exhibits a turnover as one passes from the relatively low dissipative to
the strong dissipative regime ( Eq.(87) ).

In conclusion, we have thus discussed a number of basic issues in the 
classical theory of motion of a particle in a force field in presence of 
{\it external, adiabatic} fluctuations. 
In view of several experimental
investigations on external noise-induced processes in the past, the
study of thermodynamically open systems has been specially relevant in
various areas of physical and chemical sciences. We particularly mention the
following examples : A dye laser with fluctuating pump parameter shows 
interesting qualitative changes in the stationary distributions \cite{roy}. 
The shape of the
distribution also changes as a function of increasing fluctuation strength
in the case of a two-species chemical reaction with a fluctuating rate
coefficient \cite{arnold,sancho}. The early work on electronic parametric 
oscillator driven by
external noise which exhibits the transition from non-oscillatory to
oscillatory behavior is also worth-mentioning \cite{kawa}. Although the 
driving noise
processes in the above mentioned cases are fast, suitable extension to
adiabatic noise limit (such a typical case had been discussed by us earlier
in \cite{jpa1} in detail in connection with population inversion in a 
two-level atom by adiabatically varying the field strength 
\cite{gris1,gris2,gris3,tla})
might lead to experimental situations which are
relevant to the present theoretical context. We believe that studies in
this direction are worth-pursuing.

\acknowledgments
SKB is indebted to Council of Scientific and Industrial Research (C.S.I.R.),
Government of India for a fellowship.
Partial financial support from the Department of Science and Technology,
Government of India is gratefully acknowledged.
We express our sincerest thanks to Professor J. K. Bhattacharjee 
(Department of Theoretical Physics, IACS) for various discussions during the 
progress of this work. SKB expresses his thanks to Mr. P. Chaudhury 
(Department of Physical Chemistry, IACS) for helpful discussions.


\begin{appendix}

\section{Simplifications of the terms in Eq.(3) \label{app1}}

Making use of the relations (4-8) we simplify below the four terms as 
appeared in Eq.(3).

\noindent
First term :
\begin{eqnarray}
-\nabla \cdot F_0 \; P(x,t) & = & -\frac{\partial}{\partial x} [-\frac{1}{\beta}
\; V'(x) \; P(x,t) ]  \nonumber \\
& = & \frac{1}{\beta} \frac{\partial}{\partial x} [ V'(x) \; P(x,t) ] \; \; .
\end{eqnarray}

\noindent
Second term :
\begin{eqnarray}
& & \alpha^2 \nabla \cdot \int_0^\infty 
\left \{ \langle F_1 \nabla_{-\tau} \cdot
F_1(x^{-\tau}) \rangle \left |\frac{d x^{-\tau}}{dx} \right |
d\tau \right \} \; P(x,t)  \nonumber \\
& = & \alpha^2 \frac{\partial}{\partial x} \int_0^\infty
\langle \frac{1}{\beta} \xi(t) \left [ 1 - \frac{1}{\beta} V''
(x^{-\tau})\tau \right ] \frac{\partial}{\partial x} \frac{1}{\beta}\;
\xi (t-\tau) \rangle  \left [ 1+\frac{1}{\beta} 
\; V''(x)\tau \right ] \; d\tau \; P(x,t)  \nonumber \\
& = & \alpha^2 \; \frac{1}{\beta^2} \; \frac{\partial^2}{\partial x^2}
\int_0^\infty \langle \xi(t) \; \xi(t-\tau) \rangle \; [1-{\cal O}(\tau^2)]
\; d\tau \; P(x,t) \nonumber \\
& = & \alpha^2 \; \frac{1}{\beta^2} \; \frac{\partial^2}{\partial x^2}
\int_0^\infty \langle \xi(t)\; \xi(t-\tau) \rangle 
\; d\tau \; P(x,t) \nonumber \\
& = &  \alpha^2 \; c_0 \frac{1}{\beta^2} \frac{\partial^2 P}{\partial x^2} \; \; .
\end{eqnarray}

\noindent
Third term :
\begin{eqnarray}
& & -\alpha^2 \nabla \cdot \int_0^\infty \left \{ \tau \; 
\langle F_1 \nabla_{-\tau} \cdot {\dot{F}}_1(x^{-\tau}) \rangle 
\left |\frac{d x^{-\tau}}{dx} \right | d\tau \right \} P(x,t) \nonumber \\
& = & -\alpha^2 \frac{\partial}{\partial x} \int_0^\infty \; \tau
\langle \frac{1}{\beta} \xi(t) \left [ 1 - \frac{1}{\beta} V''
(x^{-\tau})\tau \right ] \frac{\partial}{\partial x} \frac{1}{\beta}\;
\left. \frac{d\xi(t)}{dt} \right |_{(t-\tau) }
\rangle  \left [ 1+\frac{1}{\beta} 
\; V''(x)\tau \right ] \; d\tau \; P(x,t)  \nonumber \\
& \simeq & -\alpha^2 \; c_1 \frac{1}{\beta^2} 
\frac{\partial^2 P}{\partial x^2} \; \; .
\end{eqnarray}

\noindent
Fourth term :
\begin{eqnarray}
& & \alpha^2 \nabla \cdot \int_0^\infty  \left \{ \tau \; 
\langle F_1 \nabla_{-\tau} \cdot F_1(x^{-\tau}) 
\nabla_{-\tau} \cdot F_0 (x^{-\tau}) \rangle 
\left |\frac{d x^{-\tau}}{dx} \right | d\tau \right \} P(x,t)  \nonumber \\
& = & \alpha^2 \frac{\partial}{\partial x} \int_0^\infty \; \tau
\langle \frac{1}{\beta} \xi(t) \left [ 1 - \frac{1}{\beta} V''
(x^{-\tau})\tau \right ] \frac{\partial}{\partial x} \frac{1}{\beta}\;
\xi (t-\tau) \; \left [ 1-\frac{1}{\beta} V''(x^{-\tau})\tau \right ] 
\frac{\partial}{\partial x} \nonumber \\
& & \times  \left ( -\frac{1}{\beta} \right ) \;
\left [ V'(x) - \tau V''(x) \right ] \rangle  \left [ 1+\frac{1}{\beta} 
\; V''(x)\tau \right ] \; d\tau \; P(x,t)  \nonumber \\
& = & - \alpha^2 \; \frac{1}{\beta^3} \; \frac{\partial^3}{\partial x^3}
\int_0^\infty \tau \langle \xi(t) \; \xi(t-\tau) \rangle 
\; [V'(x) -\tau V''(x) ] \; [1+ \frac{1}{\beta} V''(x)\tau ] 
\; d\tau \; P(x,t) \nonumber \\
& = & - \alpha^2 \; c_2 \frac{1}{\beta^3} \frac{\partial^3}{\partial x^3}
[V'(x)\; P(x,t)] \; \; .
\end{eqnarray}


\section{Calculation of various quantities for simplification of 
Eq.(32) \label{app2}}

We explicitly calculate the several quantities which appeared in the 
expression for $P(x)$ in Eq.(\ref{gsol}). These are necessary for the derivation of 
Eq.(47). To this end we first note that
\begin{equation}
K_{n+\frac{1}{2}} (\sqrt{b} x) = \sqrt{\frac{\pi}{2\; b^{1/2}}} \;
\frac{e^{-\sqrt{b}x}}{\sqrt{x}} \; \sum_{k=0}^n 
\frac{(n+k)!}{2^k \; b^{k/2} \; k! \; (n-k)! } \left ( \frac{1}{x^k} \right )
\; \; ,
\end{equation}
\begin{equation}
I_{n+\frac{1}{2}} (\sqrt{b} x) = \frac{1}{\sqrt{2\; \pi \; b^{1/2}}} \;
\frac{e^{\sqrt{b}x}}{\sqrt{x}} \; \sum_{k=0}^n 
\frac{(-1)^k \; (n+k)!}{2^k \; b^{k/2} \; k! \; (n-k)! } 
\left ( \frac{1}{x^k} \right )\; \; .
\end{equation}

\noindent
Using the expression for $f_k^n$ (Eq.(48))
\begin{eqnarray*}
f_k^n = \frac{(n+k)!}{2^k \; b^{k/2} \; k! \; (n-k)! } 
\end{eqnarray*}

\noindent
we rewrite
\begin{equation}
K_{n+\frac{1}{2}} (\sqrt{b} x) =
\sqrt{\frac{\pi}{2\; b^{1/2}}} \; \sum_{k=0}^n f_k^n \; 
\frac{e^{-\sqrt{b}x}}{x^{k+\frac{1}{2}}} \; \; ,
\end{equation}
\begin{equation}
I_{n+\frac{1}{2}} (\sqrt{b} x) =  
\frac{1}{\sqrt{2\; \pi \; b^{1/2}}} \; \sum_{k=0}^n (-1)^k \; f_k^n \; 
\frac{e^{\sqrt{b}x}}{x^{k+\frac{1}{2}}} \; \; .
\end{equation}

\noindent
Therefore we have
\begin{equation}
x^\nu \; K_\nu(\sqrt{b} x) = \sqrt{\frac{\pi}{2\; b^{1/2}}} \; 
\sum_{k=0}^n f_k^n \; e^{-\sqrt{b} x} \; x^{n-k}
\end{equation}

\noindent
and
\begin{equation}
\int^{\sqrt{b} x} x^\nu \; K_\nu (\sqrt{b} x) \; dx = 
\sqrt{\frac{\pi}{2\; b^{1/2}}} \; \sum_{k=0}^n f_k^n \; 
\int^{\sqrt{b} x} \; e^{-\sqrt{b} x} \; x^{n-k} \; dx \; \; .
\end{equation}

\noindent
Integrating successively by parts \cite{gr} $(n-k)$ times we obtain
\begin{equation}
\int^{\sqrt{b} x} x^\nu \; K_\nu (\sqrt{b} x) \; dx = -
\sqrt{\frac{\pi}{2\; b^{1/2}}} \; e^{-\sqrt{b} x} \; \sum_{k=0}^n 
\sum_{j=0}^{n-k} f_k^n \; \frac{(n-k)!}{j!} \; 
\frac{x^j}{(\sqrt{b})^{n-k-j+1}} \; \; .
\end{equation}

\noindent
Similarly we have
\begin{equation}
\int^{\sqrt{b} x} x^\nu \; I_\nu (\sqrt{b} x) \; dx = 
\frac{1}{\sqrt{2\;\pi\; b^{1/2}}} \; e^{\sqrt{b} x} \; \sum_{k=0}^n 
\sum_{j=0}^{n-k} (-1)^{j-n} \; f_k^n \; \frac{(n-k)!}{j!} \; 
\frac{x^j}{(\sqrt{b})^{n-k-j+1}} \; \; .
\end{equation}

\noindent
Hence from (B7) and (B4) we obtain
\begin{eqnarray}
& & x^{-\nu}\; I_\nu(\sqrt{b} x) \int^{\sqrt{b} x} x^\nu \; K_\nu (\sqrt{b} x) 
\; dx \nonumber \\
& = & x^{-(n+\frac{1}{2})} \; \left ( - \sqrt{\frac{\pi}{2\; b^{1/2}}} 
\right ) \left [ \; e^{-\sqrt{b} x} \; \sum_{k=0}^n 
\sum_{j=0}^{n-k} f_k^n \; \frac{(n-k)!}{j!} \; 
\frac{x^j}{(\sqrt{b})^{n-k-j+1}} \; \right ] \nonumber \\
& & \times \frac{1}{\sqrt{2\; \pi\; b^{1/2}}} \sum_{i=0}^n (-1)^i \; f_i^n \;
\frac{e^{\sqrt{b} x}}{x^{i+\frac{1}{2}}} \nonumber \\
& = & -\frac{1}{2\sqrt{b}} \sum_{i=0}^n \sum_{k=0}^n \sum_{j=0}^{n-k}
(-1)^i \; f_i^n \; f_k^n \; \frac{(n-k)!}{j!} \; 
\frac{x^{j-i-n-1}}{(\sqrt{b})^{n-k-j+1}} \; \; .
\end{eqnarray}

\noindent
Similarly,
\begin{eqnarray}
& & x^{-\nu}\; K_\nu(\sqrt{b} x) \int^{\sqrt{b} x} x^\nu \; I_\nu 
(\sqrt{b} x) \; dx \nonumber\\
& & = \frac{1}{2\sqrt{b}} \sum_{i=0}^n \sum_{k=0}^n \sum_{j=0}^{n-k}
(-1)^{j-n} \; f_i^n \; f_k^n \; \frac{(n-k)!}{j!} \; 
\frac{x^{j-i-n-1}}{(\sqrt{b})^{n-k-j+1}} \; \; .
\end{eqnarray}
 
\noindent
The expressions (B3,B4) and (B9,B10) can now be utilized in Eq.(\ref{gsol}) 
to obtain the expression for probability distribution function $P(x)$ 
( Eq.(47) ).

\end{appendix}



\begin{figure}
\caption{
A schematic plot of the Kramers' type potential $V(x)$. 
}
\label{fig1}
\end{figure}

\begin{figure}
\caption{
The normalized probability distribution function
$P_{n+\frac{1}{2}}(x)$ is plotted as a function of $x$ for $n=1$ and $n=2$
( $\beta =1.0$, $\Delta = 2.5$ and $c_2 = 0.9$ ).
}
\label{fig2}
\end{figure}

\begin{figure}
\caption{
The normalized probability distribution function
$P_{3/2}(x)$ is plotted as a function of $x$ for various values of
the third order noise strength $c_2$ ( $\beta =1.0$ and $\Delta = 2.5$ ).
}
\label{fig3}
\end{figure}

\begin{figure}
\caption{
Escape rate ${\cal K}_{3/2}$ is plotted as a function of 
the characteristic dissipation $\beta$ of the system for various values of 
$c_2$ ( $c_{01} = 7.0$, $\omega_b = 0.80$ and $\Delta = 2.5$ ).
}
\label{fig4}
\end{figure}

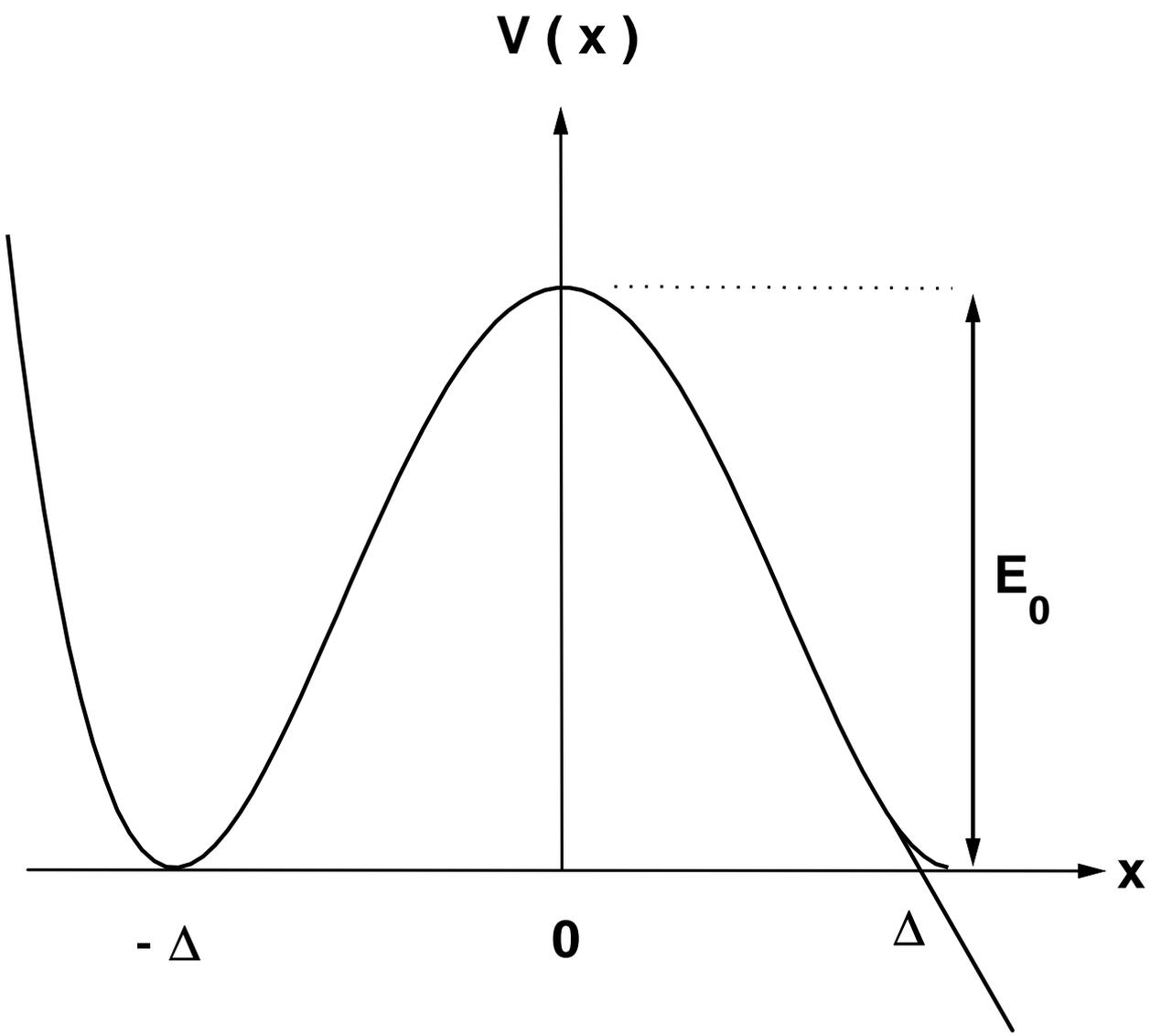
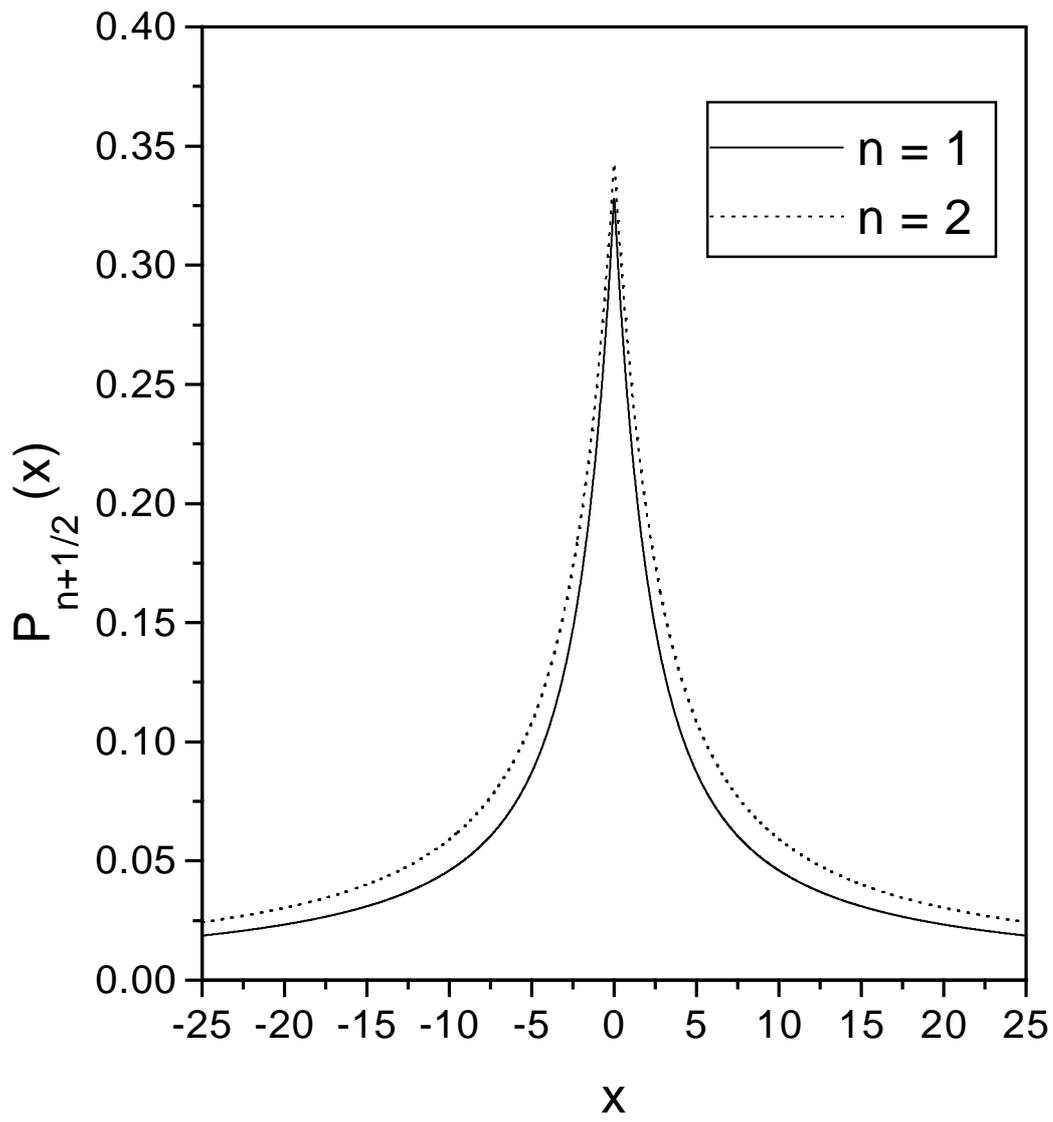
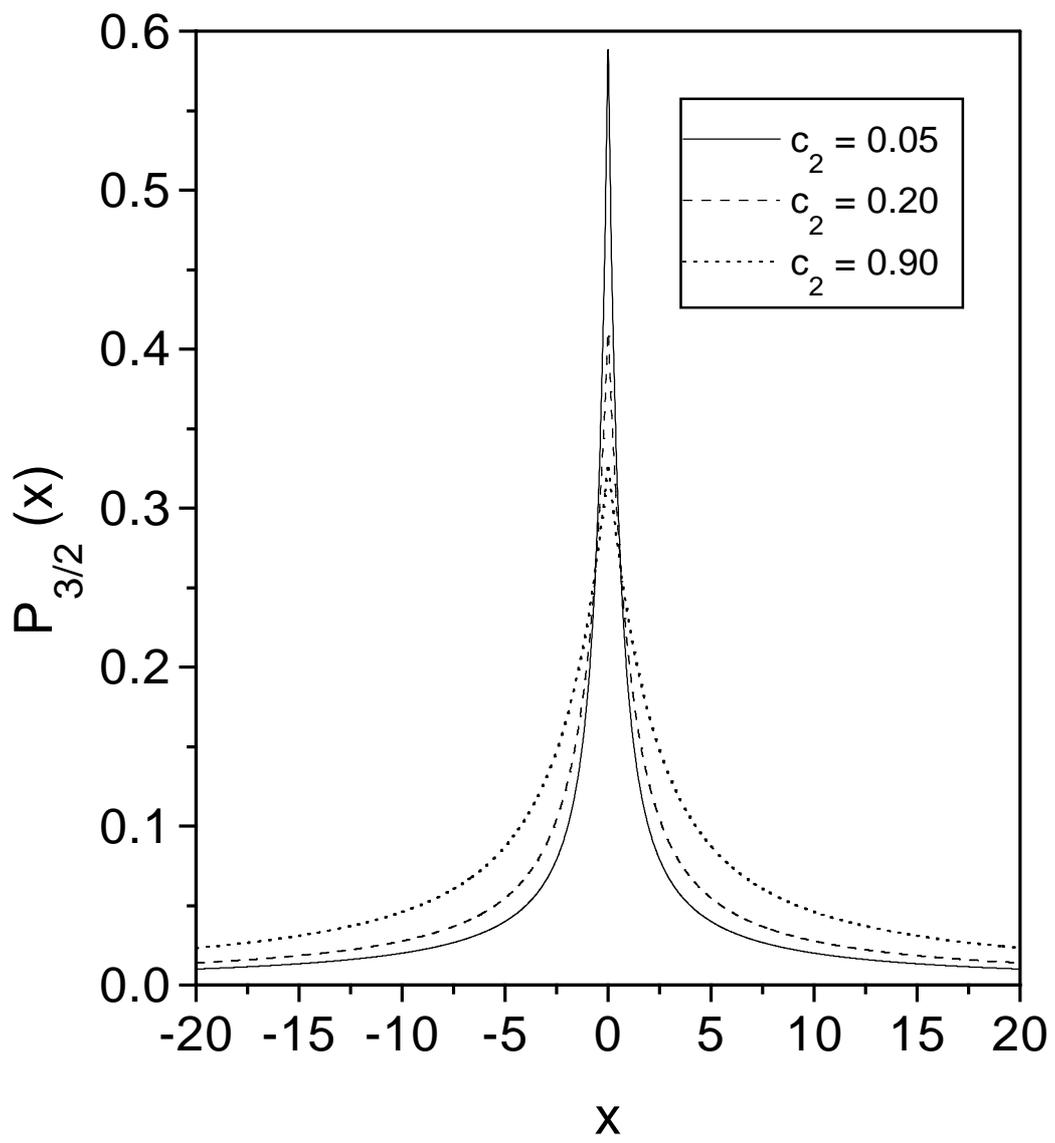
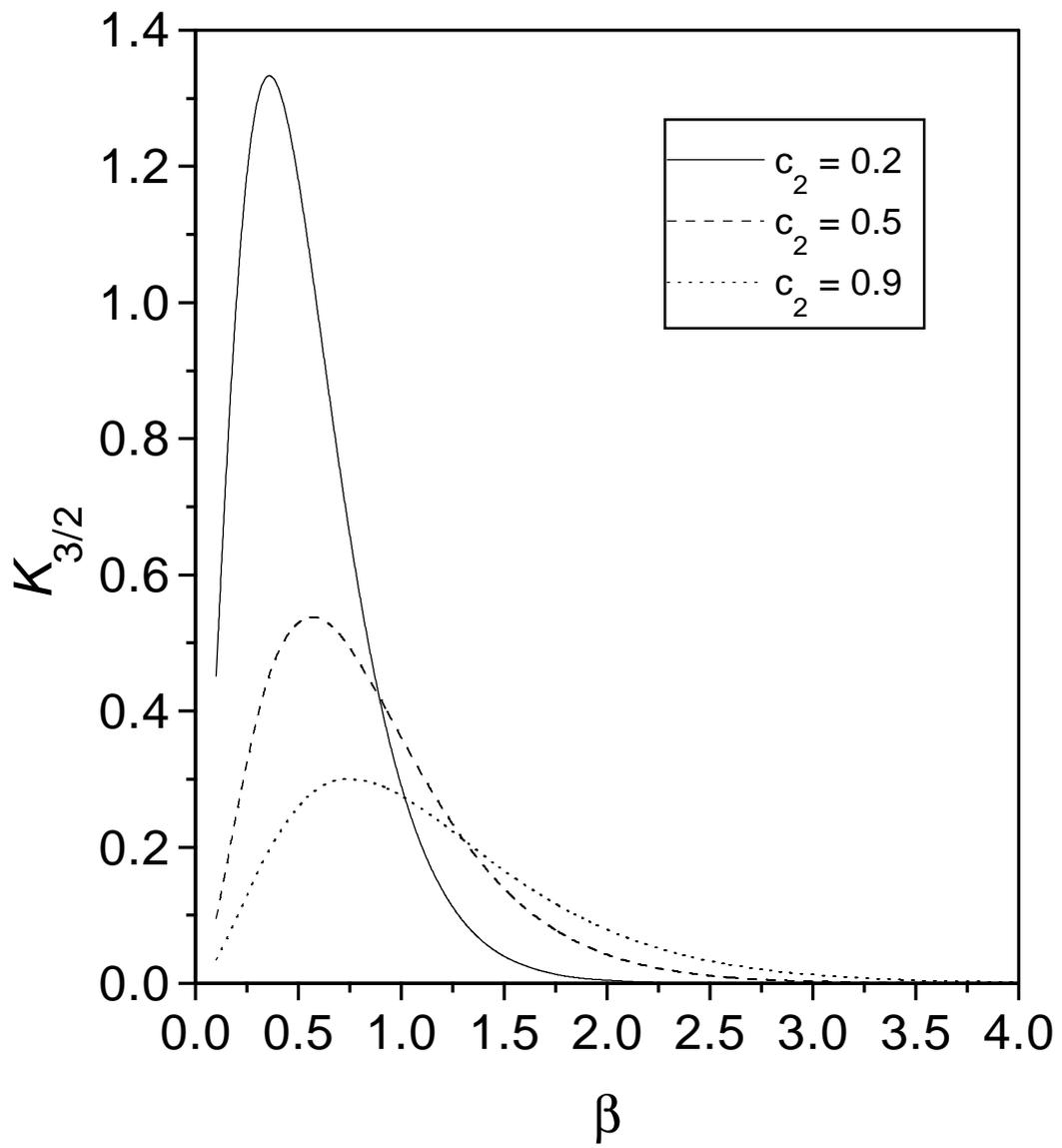
\end{document}